# All-Electronic Nanosecond-Resolved Scanning Tunneling Microscopy: Facilitating the Investigation of Single Dopant Charge Dynamics

Mohammad Rashidi[1,2], Wyatt Vine[1], Jacob A.J. Burgess[3,4,5], Marco Taucer[1,2,6], Roshan Achal[1], Jason L. Pitters[2], Sebastian Loth[3,4] and Robert A. Wolkow[1,2]

[1]Department of Physics, University of Alberta, Edmonton, Alberta, T6G 2J1, Canada.
[2]National Institute for Nanotechnology, National Research Council of Canada, Edmonton, Alberta, T6G 2M9, Canada.
[3]Max Planck Institute for the Structure and Dynamics of Matter, 22761 Hamburg, Germany.
[4]Max Planck Institute for Solid State Research, 70569 Stuttgart, Germany.
[5]Department of Physics and Astronomy, University of Manitoba, Winnipeg, Manitoba, R3T 2N2, Canada.
[6] Joint Attosecond Science Laboratory, University of Ottawa, Ottawa, Ontario, K1N 6N5, Canada.
Email Address: rashidi@ualberta.net

The miniaturization of semiconductor devices to the scales where small numbers of dopants can control device properties requires the development of new techniques capable of characterizing their dynamics. Investigating single dopants requires sub-nanometer spatial resolution which motivates the use of scanning tunneling microscopy (STM), however, conventional STM is limited to millisecond temporal resolution. Several methods have been developed to overcome this shortcoming. Among them is all-electronic time-resolved STM, which is used in this work to study dopant dynamics in silicon with nanosecond resolution. The methods presented here are widely accessible and allow for local measurement of a wide variety of dynamics at the atomic scale. A novel time-resolved scanning tunneling spectroscopy technique is presented and used to efficiently search for dynamics.

Scanning tunneling microscopy (STM) has become the premier tool in nanoscience for its ability to resolve atomic-scale topography and electronic structure. One limitation of conventional STM, however, is that its temporal resolution is restricted to the millisecond timescale because of the limited bandwidth of the current preamplifier. It has long been a goal to extend STM's temporal resolution to the scales on which atomic processes commonly occur. The pioneering work by Freeman *et. al.*[1] utilized photoconductive switches and microstrip transmission lines patterned on the sample to transmit picosecond voltage pulses to the tunnel junction. This junction-mixing technique has been used to achieve simultaneous resolutions of 1 nm and 20 ps (ref 2), but has never been widely adopted due to the requirement to use specialized sample structures. Fortunately, the fundamental insight gained from these works can be generalized to many time-resolved techniques; even though the bandwidth of the STM's circuitry is limited to several kilohertz, the non-linear I(V) response in STM allows faster dynamics to be probed by measuring the average tunnel current obtained over many pump-probe cycles. In the intervening years many approaches have been explored, the most popular of which are briefly reviewed below.

Shaken-pulse-pair-excited (SPPX) STM takes advantage of the advancements in ultrafast pulsed laser technologies to achieve sub-picosecond resolution by directly illuminating the tunnel junction and exciting carriers in the sample. Incident laser light creates free carriers that transiently enhance conduction, and modulation of the delay between the pump and probe ($t_d$) allows $dI/dt_d$ to be measured with a lock-in amplifier. Because the delay between the pump and probe is modulated, rather than the laser's intensity as in many other optical approaches, SPPX-STM avoids photoillumination-induced thermal expansion of the tip[3]. More recent extensions of this approach have extended the timescales over which SPPX-STM can be used to investigate dynamics by utilizing pulse-picking techniques to increase the range of pump-probe delay times[4]. Importantly, this recent development also provides the ability to measure $I(t_d)$ curves directly rather than via numerical integration. Recent applications of SPPX-STM have included the study of carrier recombination in single-(Mn,Fe)/GaAs(110) structures[5] and donor dynamics in GaAs[6]. Crucially, there are several limitations to this technique. First, this technique is limited to the study of semiconductors because optical pulses cannot excite free carriers in other materials. Secondly, although the tunneling current is localized to the tip, because a large area is excited by the optical pulses the signal is a convolution of the local properties and material transport. Finally, the bias at the junction is fixed at the measurement timescale so that the dynamics under study must be photoinduced.

A more recent optical technique, terahertz STM (THz-STM), couples free-space THz pulses focused on the junction to the STM tip. Unlike in SPPX-STM the coupled pulses behave as fast voltage pulses allowing for the investigation of electronically driven excitations with sub-picosecond resolution[7]. Interestingly, the rectified current generated from the THz pulses results in extreme peak current densities not accessible by conventional STM[8,9]. The technique has been used recently to study hot electrons in Si(111)-(7x7)[9] and image the vibration of a single pentacene molecule[10]. THz-pulses naturally couple to the tip, however, the necessity to integrate a THz source to an STM experiment is likely to be challenging to many experimenters. This motivates the development of other widely applicable and easily implementable techniques.

In 2010 Loth *et. al.*[11] developed an all-electronic technique where nanosecond voltage pulses applied on top of a DC offset electronically pump and probe the system[11]. The introduction of this technique offered a critical demonstration of unambiguous and practical applications of time-resolved STM to measure previously unobserved physics. Although it is not as fast as junction mixing STM, which preceded it, applying microwave pulses to the STM tip permits arbitrary samples to be investigated. This technique does not require any complicated optical methodologies or optical access to the STM junction. This makes it the easiest technique to adapt to low temperature STMs. The first demonstration of these techniques was applied to the study of spin-dynamics where a spin-polarized STM was used to measure the relaxation dynamics of spin-states excited by the pump pulses[11]. Until recently its application remained limited to magnetic adatom systems[12,13,14] but has since been extended to the study of carrier capture rate from a discrete mid-gap state[15] and charge dynamics of single arsenic dopants in silicon[15,16]. The latter study is the focus of this work.

Studies on the properties of single dopants in semiconductors have recently attracted significant attention because CMOS devices are now entering the regime where single dopants can affect device properties[17]. In addition, several studies have demonstrated that single dopants can serve as the fundamental component of future devices, for example as qubits for quantum computation[18] and quantum memory[19], and as single atom transistors[20,15]. Future devices may also incorporate other atomic-scale defects, such as the silicon dangling bond (DB) which can be patterned with atomic precision with STM lithography[21]. To this end DBs have been proposed as charge qubits[22], as quantum dots for quantum cellular automata architectures[23,24], atomic wires[25,26], and have been patterned to create quantum Hamiltonian logic gates[27] and artificial molecules[28,29]. Moving forward, devices may incorporate both single dopants and DBs. This is an attractive strategy because DBs are surface defects that can easily be characterized with STM and used as a handle to characterize single dopant devices. As an example of this strategy, DBs are used in this work as charge sensors to infer the charging dynamics of near-surface dopants. These dynamics are captured with the use of an all-electronic approach to TR-STM that is adapted from the techniques developed by Loth *et. al.*[11]

Measurements are performed on selected DBs on a hydrogen terminated Si(100)-(2x1) surface. A dopant depletion region extending approximately 60 nm below the surface, created via thermal treatment of the crystal[30], decouples the DB and the few remaining near-surface dopants from the bulk bands. STM studies of DBs have found that their conductance is dependent on global sample parameters, such as the concentration of dopants and the temperature, but individual DBs also show strong variations depending on their local environment[16]. During an STM measurement over a single DB the current flow is governed by the rate at which electrons can tunnel from the bulk to the DB ($\Gamma_{bulk}$) and from the DB to the tip ($\Gamma_{tip}$) (**Figure 1**). However, because the conduction of the DB is sensitive to its local environment the charge state of nearby dopants influences $\Gamma_{bulk}$ (**Figure 1B**), which can be inferred by monitoring the DB's conductance. As a result, the conductance of a DB can be used to sense the charge states of nearby dopants, and can be used to determine the rates at which the dopants are supplied electrons from the bulk ($\Gamma_{LH}$) and lose them to the STM tip ($\Gamma_{HL}$). To resolve these dynamics TR-STS is performed around the threshold voltages ($V_{thr}$) at which the tip induces ionization of near-surface dopants. The role of the pump and probe pulses is the same in the three time-resolved experimental techniques presented here. The pump transiently brings the bias level from below to above $V_{thr}$, which induces dopant ionization. This increases the conductance of the DB, which is sampled by the probe pulse which follows at a lower bias.

The techniques described in this paper will benefit those wishing to characterize dynamics occurring on the millisecond to nanosecond timescale with STM. While these techniques are not limited to studying charge dynamics it's crucial that the dynamics are manifested via transient changes in the conductance of states that can be probed by STM (*i.e.* states on or near the surface). Because the experimental modifications of commercially available STM systems required to perform the techniques described in this paper are modest it's anticipated these techniques will be extremely accessible to the community.

**PROTOCOL:**

Initial setup of microscope and experiments
- 1.1. Begin with a cryogenic-capable STM and associated control software.
- 1.2. Perform all experiments in ultrahigh vacuum with a base pressure < $10^{-10}$ Torr. Samples should be prepared in room temperature chambers connected directly to the STM. The STM should be cooled to cryogenic temperatures; this is especially important when investigating the charge dynamics of dopants which are thermally activated at modest temperatures.
- 1.3. Ensure the STM tip is equipped with high-frequency wiring (~500 MHz).
- 1.4. To prepare the cycles of voltage pulse pairs used for pump-probe experiments connect an arbitrary function generator with at least two channels to the tip (Figure 2). The voltage pulses should be generated independently before being summed and fed into the tip.
- 1.5. Apply the DC bias voltage used for imaging and conventional spectroscopy to the sample. In this configuration the tip will be grounded during STM imaging and conventional spectroscopy. Two radio frequency switches connected to the arbitrary function generator's output channels can be used to selectively ground the tip or connect it to the arbitrary function generator's output such that during pump-probe experiments the effective bias is $V_{sample} + V_{tip}$
- 1.6. Collect the tunneling current for all measurements through a preamplifier connected to the sample.

2. Preparation of the H-Si(100)-(2x1) Reconstruction
- 2.1. Cleave a sample from a 3–4 mΩ cm n-type arsenic doped Si(100) wafer and fix it to an STM sample holder.
- 2.2. Degas the sample by resistively heating it to 600 °C and holding it at that temperature for several hours in ultrahigh vacuum. The pressure will initially rise as the sample and sample holder degas, but should stabilize near the base pressure after several hours.
- 2.3. Degas a tungsten filament in the same chamber as the sample by heating the filament to 1800 °C and waiting for the pressure to recover.
- 2.4. Remove the oxide from the sample surface by 'flashing' (high ramp and cool rates, on the order of 30 °C/s) the sample to 900 °C and holding it at that temperature for several seconds before cooling it to room temperature. The pressure will increase several orders of magnitude from the base pressure during the flashing procedure. After all flashes throughout this procedure wait for the sample to cool and the pressure to recover to before continuing.
- 2.5. Progressively flash the sample to higher temperatures attempting to reach a final flash of 1250 °C. Record the voltage/current used to achieve this temperature. Abort any flash where the pressure rises above $9\times10^{-10}$ Torr in order to prevent the sample surface from getting contaminated. On the final flash determine the voltage/current required to heat the sample to 330 °C as the crystal is cooled, then let the sample cool to room temperature and the pressure recover before continuing.
- 2.6. Leak $H_2$ gas into the chamber at a pressure of $1\times10^{-6}$ Torr and heat the tungsten filament to 1800 °C. This has the effect of cracking the $H_2$ to molecular hydrogen. Hold the sample in these conditions for two minutes before flashing the sample to 1250 °C, holding it at that temperature for several seconds, and cooling it 330 °C. After one minute of exposure at 330 °C simultaneously close the $H_2$ leak valve, turn off the tungsten filament, and let the sample cool to room temperature.

2.7. Note that these high flash temperatures affect the distribution of dopants in the sample. Heating to 1250 °C has been found to induce a ~60 nm dopant depletion region near the sample surface[30].

2.8. A successfully prepared sample will have large (> 30 nm x 30 nm) terraces with relatively low numbers of defects and will demonstrate the classic Si(100)-(2x1) reconstruction, which features dimer rows running antiparallel to one another across step edges.

3. Assessing the Quality of the Pump-Probe Pulses at the Tunnel Junction

    3.1. Position the STM tip over an H-Si on the surface. These appear as the dimer rows in STM images (Figure 3B).

    3.2. Turn off the STM control loop

    3.3. Set $V_{sample}$ to -1.0 V, $V_{pump}$ to 0.5 V, $V_{probe}$ to 0.5 V, the width of the pump and probe pulses to 200 ns, and the rise/fall time of the pulses to 2.5 ns.

    3.4. Send a series of trains of pump and probe pulses where the relative delay of the pump and probe is swept from -900 ns to 900 ns.

    3.5. Plot the tunneling current as a function of the delay between the pump and probe. It will likely demonstrate strong ringing (Figure 4).

    3.6. Repeat steps 3.1 – 3.5 but increase the rise/fall times of the pulses. The ringing will decrease as the rise/fall times are increased. It's desired to eliminate ringing to provide the most accurate spectroscopic results, however, the time-resolution of these techniques is limited to the width of the pulses used. 25 ns rise times were used for this work.

4. **Time-Resolved Scanning Tunneling Spectroscopy (TR-STS)**

    4.1. Position the STM tip over a silicon DB. They appear as bright protrusions at negative tip-sample biases **(Figure 3B)**.

    4.2. Turn off the STM control loop.

    4.3. Send a train composed of only the probe pulse with a repetition rate of 25 kHz. Over a series of pulse trains, sweep the bias of the probe pulse over a range of 500 mV. This simple experiment is analogous to conventional STS where the conductance is sampled over a range of biases.

        4.3.1. The duration over which each pulse train (each with a different bias) is applied should be configured such that the resulting spectra have appropriate signal to noise ratio.

    4.4. Send a train composed of pump pulses set at a fixed bias (such that $V_{sample}$ + $V_{pump}$ > $V_{thr}$) with a repetition rate of 25 kHz.

        4.4.1. Pump pulses can have arbitrarily long durations (1 μsec is typically sufficient).

    4.5. Send a train composed of the pump pulses with the probe pulses following with a delay of 10 ns. In this experiment the probe pulse is sampling the state prepared by the pump pulse, rather than the equilibrium state sampled in conventional STS.

- 4.5.1. When displaying/evaluating the signal collected from this step subtract the signal obtained when only the pump pulse was applied.
- 4.6. Compare the probe only and the pump + probe signals. Any hysteresis in the two signals is an indication of dynamics that can be probed with time-resolved STM techniques. By keeping the range of the probe pulse fixed and coarsely scanning the DC offset (in 0.25 V steps, for example) one can efficiently map the entire energy range of the sample to identify dynamics accessible to the technique.
- 4.7. Pulse durations can be modified depending on the experiment. The width of the pump pulse needs to be longer than the rate at which the dopant is ionized such that it consistently ionizes the dopant. In general, probe durations should be of the same order as the dynamical process under study such that the maximum signal can be measured without sampling an average of the two conductance states. When searching for energies at which dynamics exist it is recommended that durations of the probe are minimized such that only one state of the system is measured to enhance hysteresis. As the relaxation time constants are found the duration of the probe pulse can be increased to improve the signal to noise ratio.

**5. Time-Resolved STM Measurements of Relaxation Dynamics**
- 5.1. Position the STM tip over a silicon DB and turn off the STM control loop.
- 5.2. Send a train composed of pump pulses set at a fixed bias (such that $V_{sample} + V_{pump} > V_{thr}$) with a repetition rate of 25 kHz.
    - 5.2.1. Pump pulses can have arbitrarily long durations (1 μsec is typically sufficient).
- 5.3. Send a train of pump and probe pulses. The probe pulses should have an amplitude smaller than the pumps and comparable to the range at which hysteresis occurs ($V_{probe} < V_{pump}$, $V_{probe} + V_{sample} \approx V_{hystersis}$). Sweep the delay between the pump and probe pulse up to several tens of μs. Subtract the signal obtained when only the pump pulse was applied.
- 5.4. If the signal obtained from the previous step is well fit by a single exponential decay function then the lifetime of the transient state prepared by the pump pulse can be extracted from the fit.

**6. Time-Resolved STM Measurements of Excitation Dynamics**
- 6.1. Send a train composed of pump pulses set at a fixed bias (such that greater than $V_{sample} + V_{pump} > V_{thr}$) with a repetition rate of 25 kHz.
    - 6.1.1. Sweep the duration of the pump pulse from several nanoseconds to several hundred nanoseconds.
- 6.2. Send a train of pump and probe pulses. The probe pulses should have an amplitude smaller than the pumps and comparable to the range at which hysteresis occurs ($V_{probe} < V_{pump}$, $V_{probe} + V_{sample} \approx V_{hystersis}$). Subtract the signal obtained when only the probe pulse was applied.
- 6.3. If the signal obtained in the previous step is exponential it indicates that the pump pulse is preparing the transient state (ionized dopant) at a rate that can be extracted from the fit.

**REPRESENTATIVE RESULTS:**
The results presented in this section of the text have been previously published[15,16].

**Figure 3** illustrates the behavior of an example selected DB with conventional STM. A conventional I(V) measurement (**Figure 3A**) clearly depicts a sharp change in the conductance of the DB at $V_{thr}$ = -2.0 V. This behavior is also observed in STM images taken at -2.1 V (**Figure 3B left panel**), -2.0 V (**middle panel**) and -1.8 V (**right panel**) where the DB has the appearance of a bright protrusion, a speckled protrusion, and a dark depression, respectively. This transition can also be observed by looking at the tunneling current collected with the tip positioned directly on top of the DB with the z-controller turned off and the bias set to $V_{thr}$ which results in two-level telegraph noise (**Figure 3C**). Crucially, in these measurements the tip of the STM was placed directly over the DB. In this arrangement, the fluctuation of the DB's charge state occurs on timescales below $e/I_T$ = 2ns, which is much faster than the switching observed. As such, the observed behavior is attributed to the charging dynamics of nearby dopants which affects the conductance of the DB. As the bias was increased, the frequency of the telegraph noise also strongly increased, such that at -2.02 V the switching behavior could no longer be resolved directly by the STM's preamplifier. This motivated the use of TR-STM.

**Figure 4** demonstrates a method that can be used to assess the quality of the pulses delivered to the junction. Ringing is observed when fast rise times are used for the pump and probe pulses because of the impedance mismatch between the STM circuitry and the tunnel junction. An autocorrelation of the pump and probe pulses can be generated by sweeping the relative delay of the probe pulse through zero-delay. A strong increase in tunneling current occurs when the pump and probe pulses overlap due to the characteristic non-linear behavior of I(V) measurements. Ringing is manifested through smaller amplitude oscillations in the tunneling current on both sides of the origin. By increasing the rise times on the pulses from 2.5 ns to 25 ns a strong suppression of the ringing is observed. The relative offset of the signals generated with each pulse rise time is a result of the fact that the pulse width is measured to include the rise times of the pulses. Therefore, pulses with 2.5 ns rise times have a greater integrated area, and therefore a greater integrated current, compared to pulses with 25 ns rise times. This highlights that quantitative comparison of TR-STM measurements should only be performed when the rise times of the pulses used are equal.

**Figure 5** demonstrates TR-STS. In these measurements, a pump pulse transiently brings the system above $V_{thr}$ and immediately after a probe pulse interrogates the conductance of the transient state. The conductance of the transient state can be mapped by subtracting the signal acquired with probe only from the signal obtained with pump+probe. When the pump+probe and probe-only signals are compared directly any hysteresis is indicative of dynamic processes that can be probed by TR-STS. By changing the value of the fixed DC offset bias the dynamics of the system that can probed by TR-STS can be efficiently identified.

      In TR-STS it is important to consider the duration of the pump and probe pulses. The pump pulse should be sufficiently long to induce a steady-state of the system (ie. pump it into the high conductance state). If the duration of the probe pulse is too long, however, then at low probe

amplitudes the conductance of the DB can relax during the measurement. In this case the probe pulse will sample both the high and low conductance states of the DB and reduce the visibility of the hysteresis. Therefore, to maximize the visibility of the hysteresis the duration of the probe pulse should be shorter than the relaxation rate of the high-conductance state.

**Figure 6** demonstrates time-resolved measurements of dopant relaxation and excitation dynamics. As in TR-STS, the sample is set to a fixed DC offset bias below $V_{thr}$ and the pump pulses transiently bring the system above $V_{thr}$. Relaxation dynamics were probed by sweeping the relative delay of the probe pulses (**Figure 6A**). Fitting a plot of the probe current as a function of the relative delay (**Figure 6B**) with a single exponential decay permitted $\Gamma_{HL}$ to be extracted. It is important to note that this rate is never observed for a single cycle, rather $\Gamma_{HL}$ is inferred from the time-averaged tunneling current which is comprised of many events. This is analogous to optical spectroscopy where the lifetime of an excited state can be determined from single measurements of an ensemble, except that in this case the lifetime of a single dopant can be characterized through an ensemble of measurements because it can be probed directly by the STM tip. It is important to note that the probe current observed in **Figure 6B** does not decay to zero, but rather to a fixed offset. This is because the pump pulse excites dynamics (observed as the millisecond-scale telegraph noise in **Figure 3C**) that do not decay within the measurement timeframe. This indicates that the conductance of the DB under study is gated by two dopants with distinct relaxation time constants. **Figure 6C** demonstrates a control experiment where the amplitude of the pump is varied from -0.25 V to -0.6 V. A change in the lifetime of the ionized state as a function of the pump amplitude would indicate that additional dynamical processes exist close in energy to $V_{thr}$. Because $\Gamma_{HL}$ is constant beyond -2.05 V, it's concluded that only the charge states of the two identified dopants are gating the conductance of the DB.

Excitation dynamics were probed by sweeping the width of the pump pulse (**Figure 6D**). $\Gamma_{LH}$ was extracted from an exponential fit of the time-averaged current as a function of the pump width (**Figure 6E**). When the pumped bias does not exceed $V_{thr}$ there is no observed dependence between the pump width and the tunneling current because the dopants remain neutral. When the pumped bias exceeds $V_{thr}$ an electron can tunnel from the dopant to the tip leaving the dopant ionized. By sweeping the width of the pump, the average rate at which the dopant is ionized is mapped. **Figure 6F** investigates the dependence of $\Gamma_{LH}$ as a function of the pump's amplitude. If a DB is gated by a single dopant $\Gamma_{LH}$ would be linear with the pump's amplitude over the entire bias range. This is expected because the dopant's ionization rate depends exponentially on the strength of the local electric field, which scales linearly with the bias applied to the tip. DB1, which is the DB studied in all prior figures, demonstrates a linear dependence between -2.1 V and -2.25 V and a step at -2.05 V. This step is further evidence that DB1 is gated by two nearby dopants. A linear dependence was observed for DB2 over the range -1.3 V to -1.6 V, indicating that a single dopant gated it. DB2 did not exhibit any dynamics beyond the millisecond timescale, and therefore was not studied with the other time-resolved techniques.

## DISCUSSION:

The variant of TR-STS in which the pump pulse is not applied is equivalent to conventional STS except that the system is being sampled at a high frequency rather than continuously. If the duration of the probe pulses are appropriate (>$\Gamma_{LH}$) the TR-STS signal acquired without the pump pulse can be multiplied by a constant proportional to the experiment's duty cycle to coincide exactly with a conventional STS measurement. This is only possible because the measurements are made without the use of a lock-in amplifier which would otherwise attenuate an unknown portion of the signal due to the low-pass filtering used. This is a significant difference between the methods employed by Loth *et. al.*[11] and those presented in this work. The use of a lock-in amplifier can be used to enhance the sensitivity of the measurements but prevents the direct comparison of TR-STS with conventional STS measurements. In systems where this sensitivity is required it is anticipated that both methods could be used in concert, with the lock-in amplifier turned off such that experimenters can efficiently search for dynamics, and the lock-in amplifier turned on for added sensitivity during the characterization of excitation and relaxation dynamics.

The main drawback of these techniques is that their temporal resolution is currently limited to several nanoseconds. This is several orders of magnitude slower than what can be achieved with junction-mixing or optical techniques. This is a consequence of signal ringing, which occurs when voltage pulse with sub-nanosecond rise times are used because of the impedance mismatch between the STM circuit and the tunnel junction[31]. Indeed, all-electronic methods have achieved time resolution as fine as 120 ps (ref 32) but have not yet been used to investigate dynamics at that resolution. An optimally designed STM would have a perfectly impedance-matched STM circuit up until the tunnel junction, which would be perfectly impedance-mismatched. This would eliminate distortion and dissipation of the pulse and would reflect the microwave power rather than deliver it across the junction. A possible strategy to eliminate the resulting ringing would be to add extra dissipation to the STM circuit so that reflected pulses would be effectively attenuated.

In this work, the simplest approach was taken *i.e.* no internal modification to the commercial STM was performed. An autocorrelation technique was used to characterize the ringing which was then minimized by simply extending the rise times of the pulses. Because the rise time of the pulses limits the time resolution this strategy cannot be used to characterize dynamic processes occurring on timescales at the limits of these techniques (several ns). In these situations, ringing can be actively suppressed by employing the techniques developed by Grosse *et. al.*[31] which involve shaping the pulses to account for the transfer function of the arbitrary function generator and the tunnel junction.

The all-electronic approach to TR-STM has many advantages over other prominent TR-STM approaches. Firstly, in comparison to junction-mixing STM it does not require any specialized sample structures. Any samples that can be scanned with conventional STM should be amenable to these techniques. Further, the all-electronic approach does not require significant modifications to the STM or the use of ultrafast optics. Indeed, the modifications to the STM circuitry required to perform these techniques are extremely modest, as commercial STMs with high-frequency cabling are available. Additionally, the dynamics probed with the all-electronic approach are purely local, as the pulses are supplied directly to the STM tip. This contrasts with SPPX-STM where the incident laser pulses can only be focused to several square microns. Finally, the all-electronic method permits the ability to accurately manipulate the biases of the pump and probe, allowing a direct comparison to standard STM measurements. This is central to several of the techniques described in this paper, and while it may be possible to implement similar pulse sequences in optical approaches to TR-STM it is experimentally difficult.

The experimental techniques presented here measure charge dynamics with atomic spatial resolution and nanosecond temporal resolution. There is a wealth of new physics to be studied with this very accessible approach. In particular, the dynamics of single electrons on single atoms are fascinating and technologically important examples. Single electron dynamics were previously studied within the limitations of conventional STM, but our technique opens the door for investigating similar processes over six additional orders of magnitude (from millisecond to nanosecond). Notably, this bridges the gap from the slow events typically observed in STM, to the fundamental processes that underlies them: the tunneling current. In STM experiments, the measured currents correspond to a flow of electrons at a rate on the order of one electron per nanosecond. Thus, all-electronic pump probe can enable time-resolved investigation of single electron charge dynamics over the entire range of time-scales characteristic of the STM.

**ACKNOWLEDGMENTS:**
We would like to thank Martin Cloutier and Mark Salomons for their technical expertise. We also thank NRC, NSERC, and AITF for financial support.

**DISCLOSURES:**
The authors declare that they have no competing financial interests.

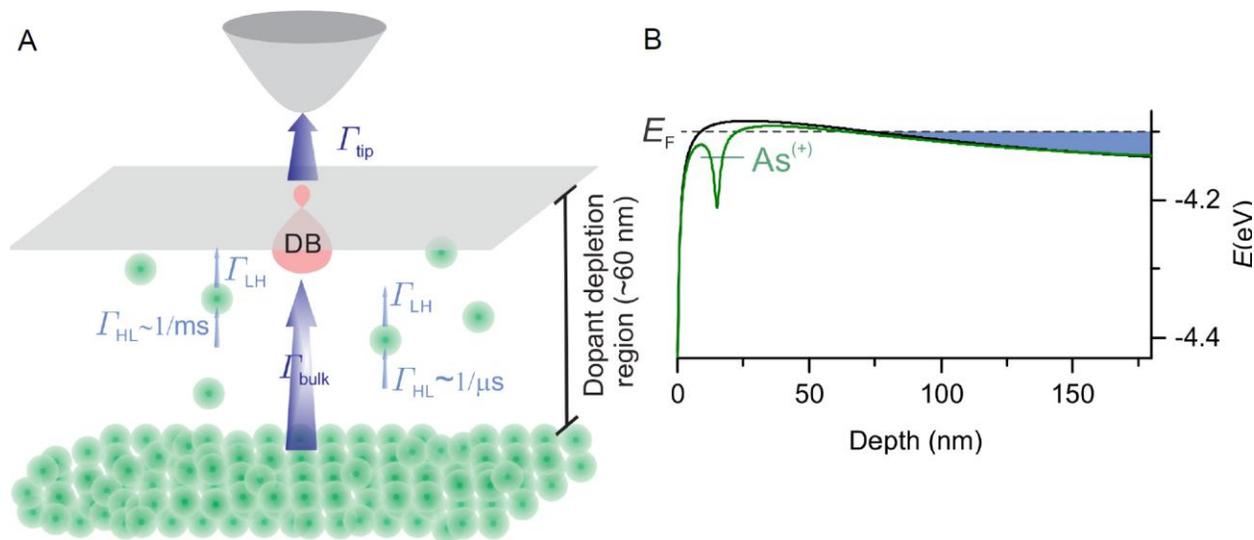

**Figure 1: Schematic of the system of study and its associated energy diagram. (A)** The current sampled by STM tip positioned directly on top of a silicon DB is predominantly composed of electrons passing from the bulk to the DB and from the DB to the tip with rates $\Gamma_{bulk}$ and $\Gamma_{tip}$, respectively. Arsenic dopants, represented by green balls, also have filling ($\Gamma_{HL}$) and emptying rates ($\Gamma_{LH}$) that can be probed by time-resolved STM measurements. **(B)** The conduction band edge in the presence of an ionized dopant (green curve) is pulled down relative to when the dopant is neutral (black curve) which results in an increased conductance. The energy diagram was computed for a sample bias of -2.0 V. The blue coloured area represents the filled states. This figure has been taken with permission from ref 16.

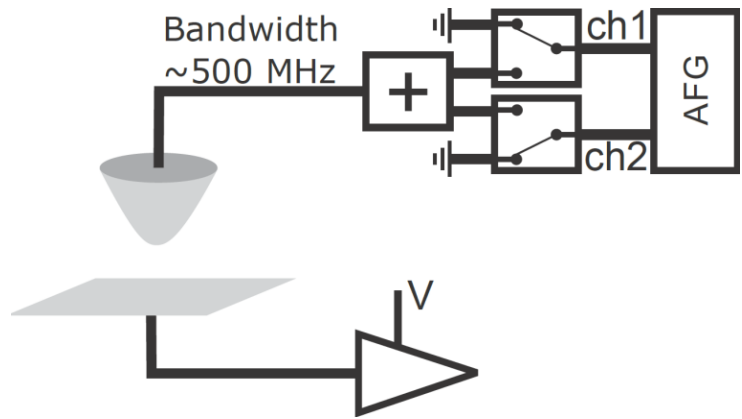

**Figure 2: The required modifications that must be made to a commercial STM such that TR-STM can be performed with the methods described in this work.** The signals created by an arbitrary function generator with two independent channels are summed and fed to the tip of an STM equipped with high-frequency cabling. Two radiofrequency switches are used to control the pulse trains. A DC offset bias is applied to the sample. The tunneling current is measured on the sample side. This figure has been taken with permission from ref 16.

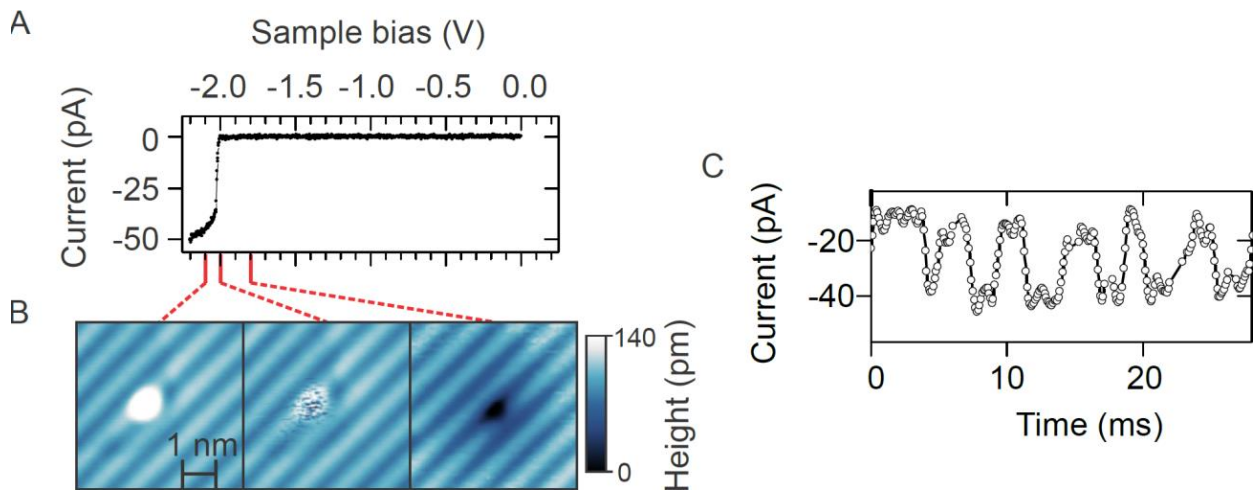

**Figure 3: The spectroscopic behavior of a selected DB with conventional STM. (A)** An I(V) measurement over the DB collected at constant height. **(B)** Constant-current STM images of the DB beyond (-2.1 V, left), at (-2.0 V, middle), and below (-1.8 V, right) the threshold voltage. **(C)** With the control loop of the STM off a current-time trace acquired over the DB at the threshold voltage (-2.01 V) displays two-state telegraph noise on the millisecond timescale. This figure has been taken with permission from ref 16.

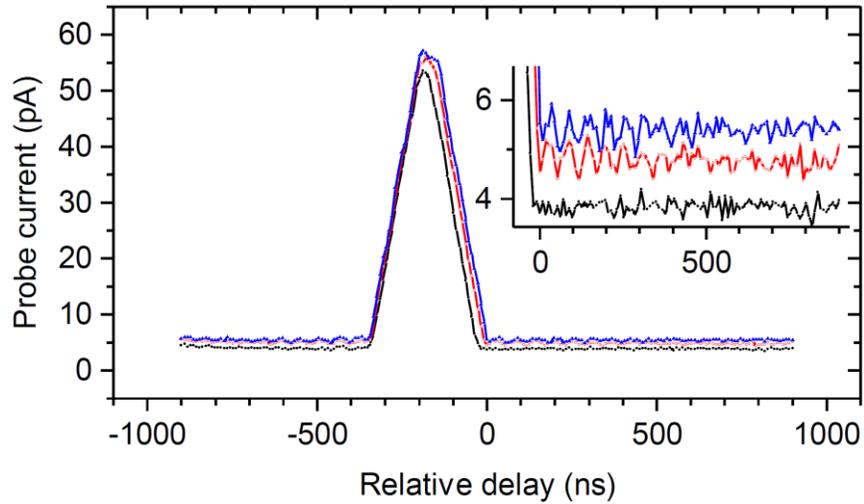

**Figure 4: An autocorrelation of the pump and probe pulses at the tunnel junction.** The tip is positioned over H-Si and a train of pump-probe pairs is delivered to the tunnel junction. The relative delay of the probe is measured from the trailing edge of the pump to the leading edge of the probe, and was swept from -900 ns to 900 ns. A static DC offset of -1.0 V was applied to the sample. Pump and probe amplitudes were set to -0.50 V with 200 ns widths. The rise/fall time of the pulses was set to 25 ns (black), 10ns (red), and 2.5 ns (blue). The probe current was multiplied by a factor of twenty to account for the 5% duty cycle employed in the measurement, however, no attempt was made to correct for the fact that the integrated areas of each pulse train differ. **Inset:** an enlarged view of the ringing between 0 and 900 ns relative delay.

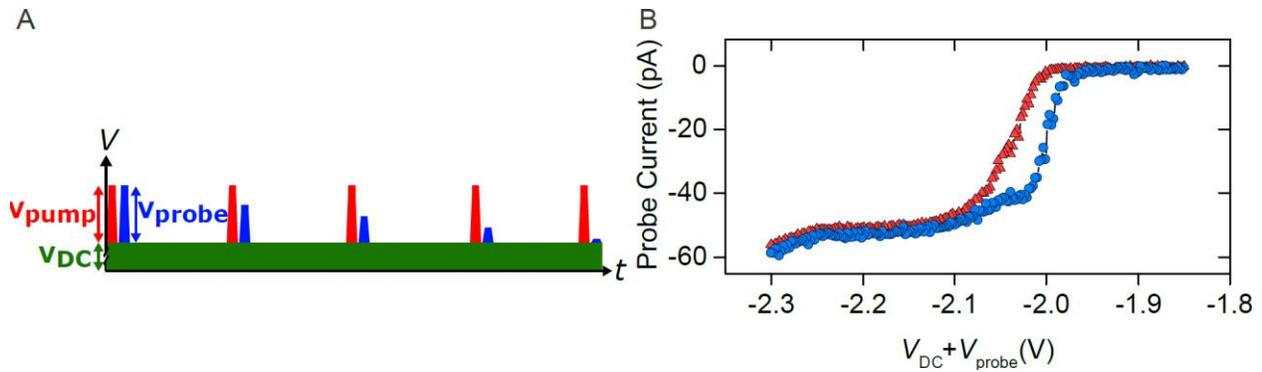

**Figure 5: Time-resolved scanning tunneling I(V) spectroscopy (TR-STS). (A)** Schematic of the TR-STS pulse sequences. A static DC offset bias (green bar) is applied to the sample. Pump pulses (red bars) precede probe pulses (blue bars). Each pair of pump and probe pulses represents a train of pulses sent to the tip. **(B)** TR-STS measurement with 1 µs width pump and probe pulses. The hysteresis between the curves without the pump (red triangles) and with the pump (blue circles) overlaps the bias range where the system is bistable. The DC bias is set to -1.80 V, the pump bias is -0.50 V, and the probe bias was swept from 500 to 50 mV. The pulses have rise/fall times of 25 ns, the relative delay between the trailing edge of the pump and the leading edge of the probe pulse is 10 ns, and the repletion rate is 25 kHz. This figure has been taken with permission from ref 16.

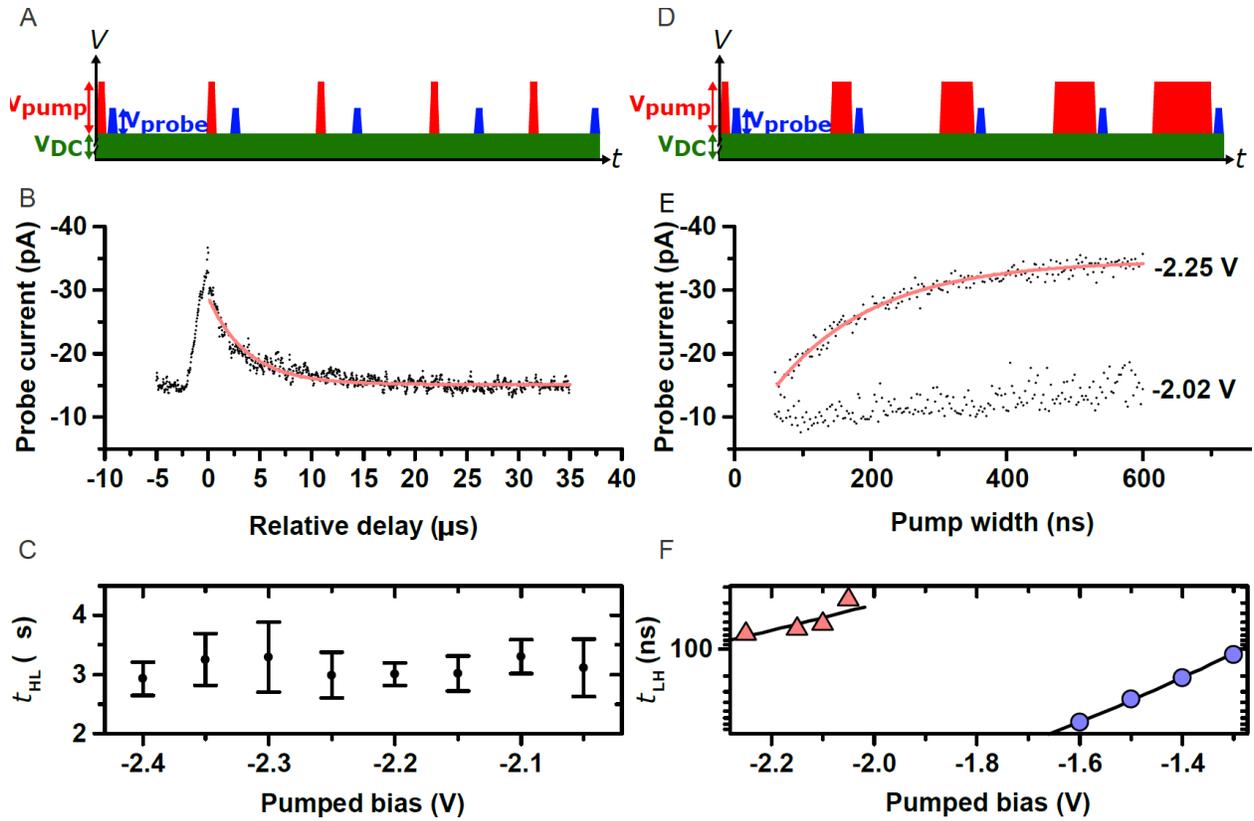

**Figure 6: Time-resolved measurements of dopant excitation and relaxation dynamics. (A, D)** Schematics of the pulse sequences. A static DC offset bias (green bar) is applied to the sample. Pump pulses (red bars) precede probe pulses (blue bars). Each pair of pump and probe pulses represents a train of pulses sent to the tip. **(B)** Measurement of $\Gamma_{HL}$ is made by sweeping the relative delay of the pump and probe pulses. The pump and probe pulses have widths of 1 µs. A DC offset of -1.80 V is applied to the sample; the pump and probe pulses have amplitudes of -0.4 and -0.21 V, respectively. **(C)** Measurements of $\Gamma_{HL}$ at different pump amplitudes. **(E)** Measurement of $\Gamma_{LH}$ is made by sweeping the duration of the pump. A DC offset of -1.80 V is applied to the sample and the probe pulses have an amplitude of -0.21 V. $V_{thr}$ for the selected DB is -2.05 V. **(F)** Measurements of $\Gamma_{LH}$ at different pump amplitudes for two selected DBs. DB1 (red triangles) is the DB used for all other measurements in the text. DB2 is a different selected DB and is described fully in ref 16. This figure has been taken with permission from ref 16.